\documentclass[twocolumn,superscriptaddress]{revtex4-2} 
\usepackage{bm}
\usepackage{changes}
\usepackage{amsmath,amssymb}
\usepackage{graphicx}

\begin{document}

\preprint{APS/123-QED}

\title{Existence and Stability of Dissipative Solitons in a Dual-Waveguide Lattice with Linear Gain and Nonlinear Losses}

\author{Zhenfen Huan}
\affiliation{Department of Physics, Changzhi University, Changzhi, Shanxi 046011, China}

\author{Changming Huang}
\email{hcm123\_2004@126.com}
\affiliation{Department of Physics, Changzhi University, Changzhi, Shanxi 046011, China}

\author{Chunyan Li}
\affiliation{School of Physics, Xidian University, Xi’an, 710071, China}

\author{Pengcheng Liu}
\affiliation{Department of Physics, Changzhi University, Changzhi, Shanxi 046011, China}

\author{Liangwei Dong}
\email{dlw\_0@163.com}
\affiliation{Department of Physics, Zhejiang University of Science and Technology, Hangzhou, China, 310023}

\date{\today}

\date{\today}

\begin{abstract}
In this study, we investigate the existence and stability of in-phase and out-of-phase dissipative solitons in a dual-waveguide lattice with linear localized gain and nonlinear losses under both focusing and defocusing nonlinearities. Numerical results reveal that both types of dissipative solitons bifurcate from the linear amplified modes, and their nonlinear propagation constant changes to a real value when nonlinearity, linear localized gain, and nonlinear losses coexist. We find that increasing the linear gain coefficient leads to an increase in the power and propagation constant of both types of dissipative solitons. For defocusing nonlinearity, in-phase solitons are stable across their entire existence region, while focusing nonlinearity confines them to a small stable region near the lower cutoff value in the propagation constant. In contrast, out-of-phase solitons have a significantly larger stable region under focusing nonlinearity compared to defocusing nonlinearity. The stability regions of both types of dissipative solitons increase with increasing nonlinear losses coefficient. Additionally, we validate the results of linear stability analysis for dissipative solitons using propagation simulations, showing perfect agreement between the two methods.
\end{abstract}

\maketitle

\section{Introduction}
\label{introduction}
The study of optical wave dynamics in nonconservative systems, especially concerning the sustained propagation of nonlinear light beams, has garnered considerable attention.
The stability of nonlinear optical waves in these systems often results from the subtle interplay between linear gain and loss mechanisms, as evidenced in $\mathcal{PT}$-symmetric systems~\cite{musslimani2008optical,ruter2010observation,konotop2016nonlinear} and general optical dissipation contexts~\cite{zezyulin2011solitons,he2012stable,he2014localized,zezyulin2016nonlinear,wang2024propagation,qiu2020soliton,mihalache2024localized}. This mechanism is further modulated by linear localized gain and nonlinear losses in two-photon absorption systems, leading to the formation of a diverse range of dissipative solitons~\cite{kartashov2010dissipative_epl,kartashov2010dissipative,kartashov2011symmetry,kartashov2019edge,kartashov2011two,huang2019dissipative,borovkova2011rotating,lobanov2011stable,borovkova2012stable,li2024stable,kartashov2010vortex,veretenov2016rotating,kominis2012gain,huang2020asymmetric,huang2020dissipative}.
Innovative research by \textit{Y.V.Kartashov} and his colleagues has expanded the understanding of these solitons, particularly in the presence of nonlinear losses. Their work encompasses the investigation of dissipative surface optical solitons at the interface between semi-infinite lattices and uniform Kerr media~\cite{kartashov2010dissipative_epl}, as well as the intriguing phenomenon of dissipative defect modes in periodic structures~\cite{kartashov2010dissipative}. Furthermore, they have explored the symmetry-breaking behavior and the emergence of multi-peak dissipative solitons~\cite{kartashov2011symmetry}, and the optical properties of edge and bulk dissipative solitons in modulated $\mathcal{PT}$-symmetric waveguide arrays~\cite{kartashov2019edge}, shedding light on the complex interplay between localized gain and nonlinear losses in one-dimensional systems. Other phenomena with high-dimensional encompass two-dimensional dissipative solitons~\cite{kartashov2011two}, surface solitons~\cite{huang2019dissipative}, vortices in uniform media~\cite{borovkova2011rotating,lobanov2011stable,borovkova2012stable,li2024stable} and in periodic lattices~\cite{kartashov2010vortex}, as well as asymmetric topological dissipative solitons~\cite{veretenov2016rotating}, have been studied.
Furthermore, studies have explored dissipative soliton routing in gain-controlled optical lattices~\cite{kominis2012gain}, asymmetric dissipative solitons in one-dimensional non-uniform gain-loss waveguides~\cite{huang2020asymmetric}, and dissipative solitons in optical chirped lattices~\cite{huang2020dissipative}.

Despite these advances, the optical properties of in-phase and out-of-phase dissipative solitons in dual-waveguide lattices have not yet been explored. In conservative systems, various intriguing optical phenomena have been observed within such lattices, including optical controlled waveguide switches~\cite{mcaulay1993designing}, optical buffers~\cite{horak2011continuously}, and optomechanical self-channeling of light~\cite{butsch2012optomechanical}. These lattices can support both in-phase and out-of-phase eigenmodes. However, the introduction of linear localized gain disrupts its self-equilibrium, favoring only the amplified modes, which is not conducive to stable long-distance light propagation. The counterbalancing role of nonlinear losses in this context is pivotal and warrants further investigation.

In this work, we construct the dual-waveguide lattice using two Gaussian-like waveguides, to investigate the existence, stability, and dynamics of in-phase and out-of-phase dissipative solitons. Both focusing and defocusing nonlinearities have been under consideration. We can find that when a balance is achieved between linear localized gain and nonlinear losses, the propagation constant of dissipative solitons bifurcating from the linear amplified mode becomes a real value. With an increase in the linear gain coefficient, both types of dissipative solitons demonstrate a consistent growth in field amplitude and propagation constant. For defocusing nonlinearity, there exists an upper cutoff in the level of linear localized gain for both types of dissipative solitons, each displaying distinct stability characteristics under focusing and defocusing nonlinearity. The stability of both types of dissipative solitons was verified through linear stability analysis and propagation simulations.
The remaining parts of this paper are organized as follows: In Section~\ref{sec2}, we introduce our theoretical model and the methods used to obtain dissipative soliton solutions. In Section~\ref{sec3}, we analyze and discuss the numerical results of linear modes and dissipative solitons in the presence and absence of nonlinearity. Finally, the paper concludes with a summary in Section~\ref{sec4}.

\section{Theoretical model}\label{sec2}
We explore the propagation of laser radiation in a dual-waveguide lattice with spatially localized linear gain and nonlinear losses, characterized by the dimensionless light field amplitude $A(x,z)$ in the following nonlinear Schr\"odinger equation:
\begin{equation}\label{eq1}
	i\frac{\partial A }{\partial z}=-\frac{1}{2} \frac{\partial^2 A}{\partial x^2} -\mathcal{V}(x)A-(\sigma+i\kappa)|A|^2A,
\end{equation}
here, $x$ and $z$ are the transverse and longitudinal coordinates, respectively. $\mathcal{V}(x) = \mathcal{V}_r(x) - i\mathcal{V}_i(x)=p_\mathrm{re}\mathcal{V}_0(x)-ip_\mathrm{im}\mathcal{V}_0(x)$ is a complex lattice, where $\mathcal{V}_r(x)$ corresponds to the refractive index distribution of the waveguide, and $\mathcal{V}_i(x)$ represents linear gain. $p_\mathrm{re}$ is the depth of the waveguide lattice, and $p_\mathrm{im}$ is the level of linear localized gain.
The function $\mathcal{V}_0(x)=\exp[-(x+x_c)^2/w^2]+\exp[-(x-x_c)^2/w^2]$, $x_c$ and $w$ respectively represent the center and width of the single waveguide. $\sigma=+1(-1)$ corresponds to focusing (defocusing) nonlinearity, and $\kappa>0$ corresponds to the level of nonlinear losses.

Stationary solutions of Eq. (\ref{eq1}) can be solved by assuming that $A(x,z)=[\phi_r(x)+i\phi_i(x)]e^{i\beta z}$, where $\phi_r$ and $\phi_i$ are the real and imaginary parts of the dissipative soliton solution, and $\beta$ is a nonlinear propagation constant.
Substituting this expression into Eq. (\ref{eq1}), we can obtain a coupled equations involving $\phi_r$ and $\phi_i$:
\begin{equation}
	\begin{array}{c}
		\frac{1}{2} \frac{\partial^{2}\phi_{r}}{\partial x^{2}}+(V_{r}-\beta) \phi_{r}+V_{i} \phi_{i}+\sigma\Xi_1-\kappa\Xi_2=0, \\
		\frac{1}{2} \frac{\partial^{2} \phi_{i}}{\partial x^{2}}+(V_{r}-\beta)\phi_{i}-V_{i} \phi_{r}+\sigma\Xi_2+\kappa\Xi_1=0.
	\end{array}
\end{equation}
Here, $\Xi_1=\phi_{r}^{3}+\phi_{i}^{2} \phi_{r}$, $\Xi_2=\phi_{r}^{2} \phi_{i}+\phi_{i}^{3}$, and from which soliton solutions in various forms can be solved by using a Newton iterative method with an additional condition: $\int_{-\infty}^{+\infty}\mathcal{V}_i(\phi_r^2+\phi_i^2)dx=\kappa\int_{-\infty}^{+\infty}(\phi_r^4+2\phi_r^2\phi_i^2+\phi_i^4)dx$.
In our discussion, we set $p_\mathrm{re}=5$, $x_c=0.78$, $w=x_c/2$, and vary $p_\mathrm{im}$, $\sigma$, and $\kappa$.
Also, Eq. (\ref{eq1}) conserves several quantities, including the power:
$U=\int_{-\infty}^{+\infty} \left | \phi_r(x)+i\phi_i(x) \right |^2dx$.

\section{Results and discussion}\label{sec3}
\begin{figure}[htb]
	\centering
	\includegraphics[width=0.8\columnwidth]{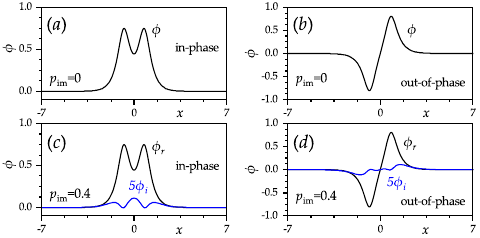}
	\caption{In-phase (a, c) and out-of-phase (b, d) linear modes. The imaginary part vanishes in (a) and (b) with $p_\mathrm{im}=0$. The blue lines in (c) and (d) with $p_\mathrm{im}=0.4$ represent the magnified imaginary parts, enhanced by a factor of $5$ for improved visibility.}
	\label{fig1}
\end{figure}

Before studying the properties of dissipative soliton solutions, it is highly instructive to investigate the characteristics of linear modes in the linear system.
Setting $\sigma=0$ and $\kappa=0$ and substituting
$A(x,z)=\phi(x)e^{i\beta z}$ into Eq.~(\ref{eq1}), we can obtain:
$\beta\phi(x)=\frac{1}{2}\frac{\partial^2\phi(x)}{\partial x^2}+\mathcal{V}(x)\phi(x)$.
Utilizing the Fourier collocation method~\cite{yang2010nonlinear}, one can extract the linear eigenvalues and modes of this equation. In the scenario of the dual-waveguide lattice with $p_\mathrm{im}=0$, the first and second eigenvalues are 2.377 and 1.875, respectively, representing the in-phase and out-of-phase modes illustrated in Figs.~\ref{fig1}(a) and \ref{fig1}(b). These two linear eigenmodes are intrinsic and robust.
For $p_\mathrm{im}=0.4$, the first and second largest eigenvalues are 2.375-0.237\textit{i} and 1.873-0.257\textit{i}, respectively.
In this case, both the in-phase and out-of-phase modes exhibit imaginary components.
This indicates that introducing linear localized gain transforms the linear modes into linear gain modes, as shown in Figs.~\ref{fig1}(c) and \ref{fig1}(d).

\begin{figure}[htb]
	\centering
	\includegraphics[width=0.8\columnwidth]{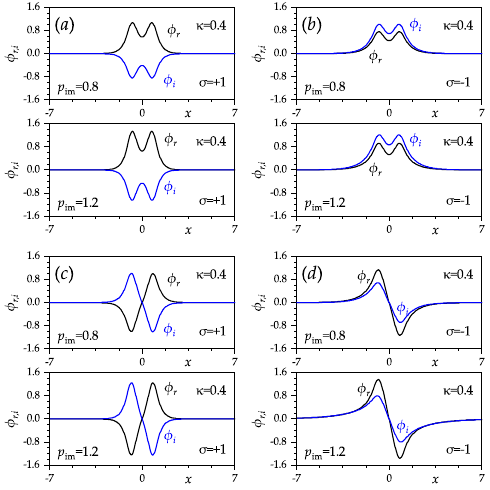}
	\caption{The profiles of dissipative solitons are depicted in-phase (a, b) and out of phase (c, d), with black lines denoting the real part and blue lines denoting the imaginary part. $p_\mathrm{im}=0.8$ in the upper panels, and $p_\mathrm{im}=1.2$ in the lower panels. $\sigma=1$ in (a) and (c), while $\sigma=-1$ in (b) and (d). $\kappa=0.4$ in all panels.}
	\label{fig2}
\end{figure}

Within the dual waveguide structure, our focusing is drawn to the existence of dissipative solitons. Two distinct types of dissipative solitons are identified within this configuration, as depicted in Fig.~\ref{fig2}. Solitons featuring matching peak values in the real components of both waveguides are denoted as in-phase dissipative solitons, whereas those with opposing peak values in the real components of both waveguides are labeled as out-of-phase dissipative solitons.
As $p_\mathrm{im}$ increases, both types of soliton profiles experience a gradual increase in amplitudes. Under the influence of focusing nonlinearity ($\sigma=+1$), the peaks of dissipative solitons' real and imaginary parts exhibit opposite trends in their corresponding waveguides [see Figs.~\ref{fig2}(a) and \ref{fig2}(c)]; conversely, with the presence of defocusing nonlinearity ($\sigma=-1$), the peaks of dissipative solitons' real and imaginary parts align within their respective waveguides [see Figs.~\ref{fig2}(b) and \ref{fig2}(d)].

\begin{figure}[htb]
	\centering
	\includegraphics[width=0.8\columnwidth]{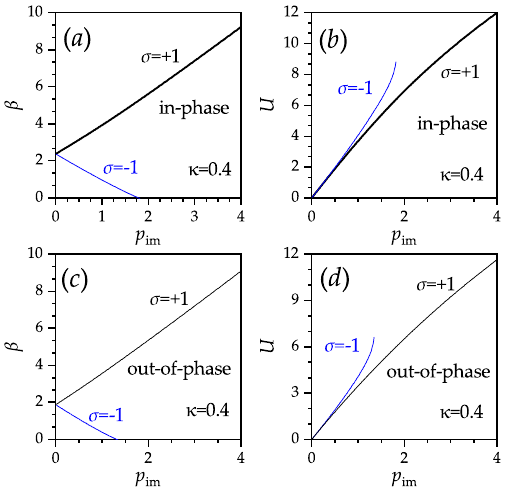}
	\caption{The propagation constant $\beta$ (a,c) and power $U$ (b,d) of in-phase and out-of-phase dissipative solitons exhibit variations with parameter $p_\mathrm{im}$. The black line denotes focusing nonlinearity ($\sigma =+1$), whereas the blue line signifies defocusing nonlinearity ($\sigma =-1$).}
	\label{fig3}
\end{figure}
At a fixed level of nonlinear loss ($\kappa=0.4$), the propagation constant $\beta$ of dissipative solitons exhibits a gradual increase with $p_\mathrm{im}$ for $\sigma=+1$ and a decrease with $p_\mathrm{im}$ for $\sigma=-1$ as illustrated in Figs.~\ref{fig3}(a) and \ref{fig3}(c). For $\sigma=+1$, as $p_\mathrm{im}$ approaches 0, the propagation constants $\beta$ converge to 2.377 for in-phase dissipative solitons and 1.875 for out-of-phase dissipative solitons, indicating the bifurcation of in-phase and out-of-phase dissipative solitons from their respective linear modes. The $U(p_\mathrm{im})$ curves for these solitons, under both types of nonlinearity, display monotonically increasing trends, with a distinct cutoff at $p_\mathrm{im}$ observed only under defocusing nonlinearity as depicted in Figs.~\ref{fig3}(b) and \ref{fig3}(d).

The investigation in this study focuses on the stability of dissipative optical solitons bifurcating from linear modes in the presence of compensating linear localized gain with nonlinear losses. The perturbed form of the dissipative soliton, governed by Eq. (\ref{eq1}), is represented as $A(x,z)=[\phi_r+i\phi_i+(u+iv)e^{\delta z}]e^{i\beta z}$, where $\delta=\delta_r+i\delta_i$, and $u$ and $v$ denote the real and imaginary components of the perturbation. By substituting this expression into Eq. (\ref{eq1}) and subsequently linearizing it, we derive the coupled equations as follows:
\begin{eqnarray}
	\delta u & = & \{ -\frac{1}{2}  \frac{\partial^2 }{\partial x^2}-V_r+\beta -[\sigma (3\phi_i^2+\phi_r^2)+2\kappa \phi _r\phi _i] \}v \nonumber\\
	& &-[2\sigma\phi _r\phi _i+\kappa (3\phi_r^2+\phi_i^2)-V_i]u, \\
	\delta v & = &  \{+\frac{1}{2}  \frac{\partial^2 }{\partial x^2}+V_r-\beta +[\sigma (3\phi_r^2+\phi_i^2)-2\kappa \phi _r\phi _i] \}u \nonumber\\
	& &+[2\sigma\phi _r\phi _i-\kappa (3\phi_i^2+\phi_r^2)+V_i]v.
\end{eqnarray}
These equations can be solved numerically using the eigenvalue package. When the maximum eigenvalue $\delta_r>0$, the dissipative soliton is unstable; otherwise, it is stable.

\begin{figure}[htb]
	\centering
	\includegraphics[width=0.8\columnwidth]{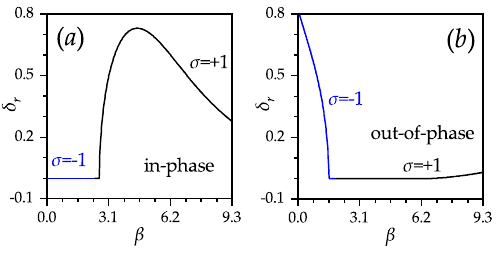}
	\caption{The relationship between the maximum unstable growth rate $\delta_r$ of in-phase (a) and out-of-phase (b) dissipative solitons and the propagation constant $\beta$ at $\kappa=0.4$.
		The blue line corresponds to defocusing nonlinearity ($\sigma=-1$), while the black line corresponds to focusing nonlinearity ($\sigma=+1$).}
	\label{fig4}
\end{figure}

\begin{figure}[htb]
	\centering
	\includegraphics[width=0.8\columnwidth]{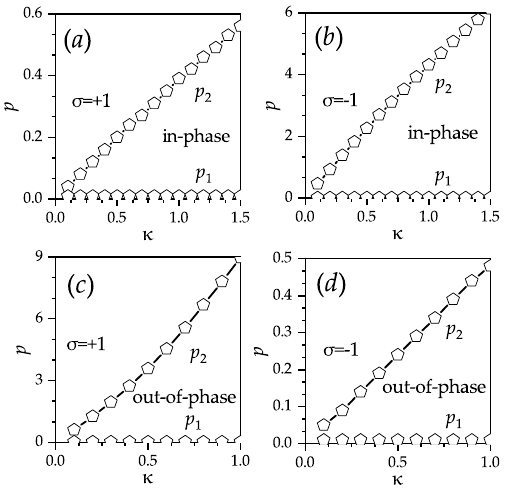}
	\caption{The stable regions $p_\mathrm{im}\in[p_1,p_2]$ of in-phase (a, b) and out-of-phase (c, d) dissipative solitons. Panels (a) and (c) correspond to $\sigma =+1$, while panels (b) and (d) correspond to $\sigma =-1$.}
	\label{fig5}
\end{figure}
\begin{figure*}[htbp]
	\centering
	\includegraphics[width=1.2\columnwidth]{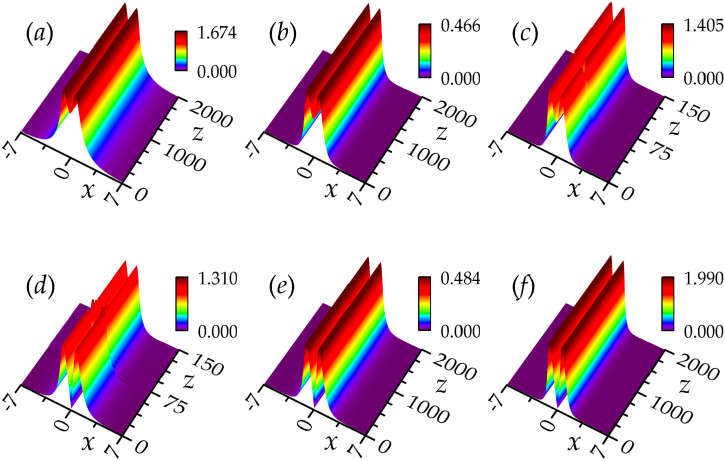}
	\caption{Examples of the propagation of stable (a, b, e and f) and unstable (c, d) dissipative solitary waves, both in-phase (a-c) and out-of-phase (d-f).
		$p_\mathrm{im}=1.5$, $\sigma=-1$ in (a),
		$p_\mathrm{im}=0.1$, $\sigma=+1$ in (b),
		$p_\mathrm{im}=0.6$, $\sigma=+1$ in (c),
		$p_\mathrm{im}=0.6$, $\sigma=-1$ in (d),
		$p_\mathrm{im}=0.1$, $\sigma=-1$ in (e) and
		$p_\mathrm{im}=1.5$, $\sigma=+1$ in (f).
	}
	\label{fig6}
\end{figure*}

Figure~\ref{fig4} illustrates the stability of in-phase and out-of-phase dissipative solitons. In-phase solitons are stable within their existence domain under defocusing nonlinearity. However, under focusing nonlinearity, stable in-phase solitons are only present for smaller values of $p_\mathrm{im}$ or $\beta$. Once $p_\mathrm{im}$ (or $\beta$) exceeds a certain threshold, in-phase solitons become unstable [see Fig.~\ref{fig4}(a)]. On the other hand, out-of-phase dissipative solitons have a larger stable region near the lower cutoff values for focusing nonlinearity, while under defocusing nonlinearity, they exhibit a narrower stable region at smaller $p_\mathrm{im}$ values [see Fig.~\ref{fig4}(b)].

Furtheremore, we conducted a systematic investigation into the existence and stability of two types of dissipative solitons under different levels of nonlinear losses, as illustrated in Figs.~\ref{fig4} and \ref{fig5}. Dissipative solitons with focusing nonlinearity do not have an upper limit, whereas those with defocusing nonlinearity exist within a specific range of linear localized gain $p_{im} \in [{p}'_1,{p}'_2]$. The stability analysis of in-phase and out-of-phase dissipative solitons within the dual-waveguide lattice yields the following insights: (i) For $\sigma=\pm 1$, the stable regions for both types of dissipative solitons expand with increasing nonlinear dissipation coefficient $\kappa$, as depicted in Fig.~\ref{fig5}. (ii) In the case of in-phase dissipative solitons with $\sigma=-1$, their existence region broadens, with all solitons within this region demonstrating stability. However, the stable region for in-phase dissipative solitons with $\sigma=+1$ is approximately one-tenth of that observed for $\sigma=-1$, as shown in Figs.~\ref{fig5}(a) and \ref{fig5}(b). (iii) In comparison to the stability of in-phase dissipative solitons, the stable region for out-of-phase solitary waves at $\sigma=+1$ is significantly larger than that at $\sigma=-1$, as illustrated in Figs.~\ref{fig5}(c) and \ref{fig5}(d).

The stability of dissipative solitons was further verified through propagation simulations for both types. The initial input beam consisted of stationary dissipative soliton solutions, represented as $A(x,z=0)=\phi_r(x)+i\phi_i(x)$. Figures \ref{fig6}(a-c) depict the typical evolutions of stable and unstable in-phase dissipative solitons. The field profiles of stable in-phase dissipative solitons remained consistent over long propagation distances, while unstable solitons fragmented after shorter distances. However, in the presence of linear localized gain and nonlinear loss in the dual-waveguide lattice, the optical wave could self-modify, producing novel field profiles. Both stable and unstable out-of-phase dissipative solitons exhibited patterns resembling those of in-phase solitons [see Figs.~\ref{fig6}(d-f)]. The propagation results for both types of dissipative solitons aligned entirely with the linear stability analyses.

\section{Conclusion}\label{sec4}
In this work, we investigated the optical properties of dissipative solitons in a one-dimensional dual-waveguide lattice, including their existence and stability. In the presence of linear localized gain, in-phase and out-of-phase dissipative solitons can bifurcate from their respective amplified modes. Under appropriate parameters, a balance between nonlinear losses and linear localized gain can be achieved, allowing stable dissipative solitons to be obtained in this structure. In-phase and out-of-phase dissipative solitons exhibit real and imaginary parts with opposite signs within their respective waveguides under focusing nonlinearity, while under defocusing nonlinearity, the real and imaginary parts have the same sign within their respective waveguides. These differences in the soliton profile distributions lead to distinct stabilities of the two types of solitons under the two nonlinearities. We found that in-phase dissipative solitons are stable across their entire existence region under defocusing nonlinearity, while under focusing nonlinearity, they exist only in a narrow stable region near the lower cutoff value in the propagation constant. In contrast, out-of-phase solitons exhibit stable regions near the upper and lower cutoff values in the propagation constant under defocusing and focusing nonlinearities, respectively. The stability regions of both types of dissipative solitons increase with increasing linear gain coefficient. Interestingly, stable dissipative solitons can propagate over considerable distances, while unstable dissipative solitons experience a breakdown in the field distribution after a short propagation distance, but quickly self-modify. These intriguing optical phenomena provide new insights into the physical characteristics of in-phase and out-of-phase solitons in dissipative systems.

\section*{Declaration of Competing Interest}
The authors declare that they have no known competing financial interests or personal relationships that could have appeared to influence the work reported in this paper.

\section*{Acknowledgements}
This work was supported by the Applied Basic Research Program of Shanxi Province (202303021211191); Scientific and Technologial Innovation Programs of Higher Education Institutions in Shanxi (2022L509).

\section*{Data availability}
Data underlying the results presented in the article are not publicly available at this time but may be obtained from the authors upon reasonable request.

\bibliographystyle{unsrt}

\end{document}